\documentclass{article}
\usepackage{graphicx}
\title{Renormalization Group Therapy}

\author{E.T. Tomboulis\footnote{e-mail: tombouli@physics.ucla.edu}  \ and 
A. Velytsky\footnote{e-mail: vel@physics.ucla.edu}\\
{\em\small 
Department of Physics and Astronomy, UCLA, Los Angeles, CA 90095-1547, USA}}
\begin{document}
\maketitle
\begin{abstract}
We point out a general problem with the procedures commonly used  
to obtain improved actions from MCRG decimated 
configurations. Straightforward measurement of the couplings from the 
decimated configurations, by one of the known methods, can result into 
actions that do not correctly reproduce the physics on the undecimated 
lattice. This is because  the decimated configurations are generally 
not representative of the equilibrium configurations 
of the assumed form of the effective 
action at the measured couplings. Curing this  
involves fine-tuning of the chosen  MCRG decimation procedure, which is 
also dependent on the form assumed for the effective action. 
We illustrate this in decimation studies of the SU(2) LGT using 
Swendsen and Double Smeared Blocking decimation procedures. A single-plaquette 
improved action involving five group representations and 
free of this pathology  is given.  
\end{abstract}

\pagebreak

\section{Introduction}

The construction of `improved actions' which reduce discretization errors 
and allow computation on coarser lattices 
has been a long-standing area of interest among lattice field theory workers. 
Ideally, one searches for the `perfect action' (\cite{Hasenfratz:1997ft} and 
references within),  
for which, under successive blocking transformations, the 
flow is along the Wilsonian `renormalized trajectory', and lattice artifacts 
disappear. Explicit construction, however, has proved rather cumbersome.

A more modest but more readily implementable approach is 
based on the Monte Carlo Renormalization Group (MCRG) 
which considers block spinning transformations on configurations 
obtained by MC simulations. The basic assumption here is that the 
block-spinned configurations are distributed according to the 
Boltzmann weight of some effective action that resulted from the 
blocking to the coarser lattice. 
One, however, does not know at the outset what this block effective action is. 
Now, under RG evolution any starting action generally develops a variety of 
additional couplings. An adequate model of the resulting effective 
action must, therefore, include a choice of several 
such couplings. By practical necessity, any Ansatz for such a model is 
restricted to some subclass of possible interactions. In the past 
effective models have been studied 
with actions consisting of one or more loops beyond the single plaquette
in the fundamental representation 
\cite{deForcrand:1999bi}, or a  
mixed fundamental-adjoint single plaquette
action \cite{Creutz:1984fj}, \cite{Hasenbusch:2004yq}. 
After a block spinning is performed starting from a simple
(e.g. Wilson) action, one needs to measure the set of couplings retained in 
one's model of the effective action.   
This may be achieved by the use of demon \cite{Creutz:1983ra} or
Schwinger-Dyson methods \cite{Gonzalez-Arroyo:1986ck}. 

There is a variety of issues that come up in the 
actual implementation of such a program. Any numerical decimation procedure 
entails some mutilation and possible loss 
of information encoded in the original undecimated configurations. 
The first thing to be checked then is that the adopted decimation 
prescription correctly reproduces physics at least at 
intermediate and long distance scales.

Assuming this is the case, some effective action must next be assumed. 
It should be noted that the issue of the choice of an 
effective action model  is not divorced from the choice of 
the decimation procedure. Indeed, in any exact RG transformation, 
the particular specification, in terms of the original variables, 
of the blocking procedure and block variables will affect the form 
of the action which results after integration over the original 
variables. Thus different decimation procedures, or even different regimes 
of the parameters entering the specification of one particular 
decimation procedure, may be fitted better by different 
effective action models.

Having adopted some class of effective actions, 
errors due to the truncation of the phase space 
inherent in any such choice may, of 
course, be significant, and prevent the model from 
adequately approaching the renormalized trajectory. 
This is something that one can 
in principle check by measurement of appropriate observables probing the 
scale regime(s) of interest, and may result in the need for addition 
of further effective couplings. 

There is, however, a more subtle pitfall lurking in a straightforward 
application of such methods. A straightforward `measurement' of the 
effective action couplings from the 
decimated configurations, by any one of the available 
methods, can actually lead to quite erroneous results. This is  because 
the decimated configurations will generally 
{\it not} be representative of the equilibrium configurations 
of the effective action at the measured couplings. As a result 
simulations on the coarser lattice with the effective action 
at these values of the couplings will not, in general, 
correctly reproduce the physics 
encoded in the decimated configurations obtained from the original theory. 
As far as we know,  
this does not appear to have been realized in 
previous MCRG gauge theory studies. 
In this paper we find that this 
problem is actually generally present and has to be dealt with.   
We use the demon method which provides a clear demonstration 
of the problem as it allows comparison between the 
original decimated and microcanonically-evolved decimated configurations in 
relation to equilibrium configurations of the assumed effective 
action.

Specifically, we explore these issues in numerical decimations 
in $SU(2)$ LGT. A preliminary report on some of our findings was 
previously given in \cite{Tomboulis:2006si}.
The present paper is organized as follows. 
In section 2 we introduce different decimation schemes, as well as 
the numerical methods we use to implement them. We check that 
the decimation procedures correctly reproduce long distance physics. We    
then adopt a single plaquette effective action which includes 
several (typically five to eight) successive group representations. There are 
special motivations for such actions originating in exact analytical results. 
In section 3 we present the results of our numerical study. We examine the 
relation between the configurations obtained by decimation from the 
original action and the thermalized configurations of the effective action.   
We explore how this affects the determination of the effective couplings    
and the tuning of the free parameters (such as the relative weight 
of staples) entering in the specification of the decimation scheme.   
We then examine how physics is reproduced by measuring observables 
such as Wilson loops of various sizes. The 
behavior under repeated decimation is also 
examined. In section 4 we extract from our results our improved action. 
Our conclusions and outlook are summarized  in section 5.

\section{Decimation procedures}

In our study we choose to start with the standard Wilson
action with coupling $\beta$ on the original undecimated lattice. 
Throughout this paper we use blocking $a \to ba$ with scale factor  
$b=2$ in all lattice directions. We employ two well-known numerical 
blocking procedures. In terms of the usual lattice gauge field bond 
variables $U_\mu(n)$, these are: 
\begin{itemize}
\item Swendsen decimation \cite{Swendsen:1981rb}
\begin{equation}
Q_\mu(n)=U_\mu(n)U_\mu(n+\hat{\mu})+c\sum_{\nu\neq\mu}U_\nu(n)U_\mu(n+\hat{\nu})
U_\mu(n+\hat{\nu}+\hat{\mu})U_{-\nu}(n+\hat{\nu}+2\hat{\mu}) \label{eq:sdec}
\end{equation}
\item Double Smeared Blocking (DSB)\cite{DeGrand:1994zr}
\begin{eqnarray}
U_\mu(n)&=&(1-6c)U_\mu(n)+c\sum_{\nu\neq\mu}U_\nu(n)U_\mu(n+\hat{\nu})
U^\dagger_\nu(n+\hat\mu)\quad \times 2\, {\rm times}\nonumber\\
Q_\mu(n)&=&U_\mu(n)U_\mu(n+\hat\mu). \label{eq:dsbdec}
\end{eqnarray}
\end{itemize}
Here $c$ is a parameter which controls the relative weight of 
staples.\footnote{More elaborate decimation procedures, such a combination 
of Swendsen and DSB, involving more than one adjustable parameter  may be 
defined. They allow more fine-tuning control in the construction 
of an appropriate improved action. The simpler, one-parameter decimation 
prescriptions (\ref{eq:sdec}) or (\ref{eq:dsbdec}), however, suffice for our 
purposes in this paper.} 
For Swendsen decimation the values $c=0.5$ and $1$ \cite{deForcrand:1999bi,
Swendsen:1981rb}, whereas for 
double smeared blocking the classical limit value 
$c=0.077$ \cite{Takaishi:1995ve} have previously been used. Fixing the 
parameter $c$ on a rational rather than ad-hoc basis will be one of our 
concerns below.

As a typical check on how such decimations preserve the 
information in  the original undecimated configurations, at least at 
long distances, we look at the quark potential. The original lattice
 potential was computed with high accuracy (20 independent runs
each consisting of 300 measurements) using the 
L\"uscher-Weisz procedure \cite{Luscher:2001up}, 
while the decimated potentials were computed in
the straightforward 'naive' way, which affects their accuracy, but suffices
for a check (from 80 to 160 independent runs each of 3000 measurements). 
The result of the comparison for Swendsen decimation 
is given in Table \ref{tab:dec_string}, with average 
goodness of fit $Q \sim 0.6$.
\begin{table}[ht]
 \centering
 \begin{tabular}{|c|c|c|}
   \hline
   $c$ &$\mu$ &$\sigma$ \\\hline\hline
   orig&0.308(4)  &0.0313(2)\\\hline
   0.1 &0.67(13)  &0.020(4)   \\
   0.12&0.71(12)  &0.019(4)    \\
   0.2 &0.33(7)   &0.031(2)    \\
   0.26&0.39(4)  &0.029(1)    \\
   0.3 &0.39(2)  &0.0292(7)  \\
   0.5 &0.36(2)  &0.0301(9)   \\
   1.0 &0.33(2)  &0.0314(6)   \\\hline
 \end{tabular}
\caption{\label{tab:dec_string}String tension $\sigma$ and Coulomb term 
coefficient $\mu$ computed on the original and 
decimated lattices. Swendsen decimation, $\beta=2.5$.
}
\end{table}
One can note that for a wide range of $c$ starting from value $\sim0.2$ all 
the decimations produce correct values of string tension $\sigma$. 
The Coulomb term coefficient $\mu$, however, which is representative of short 
distance physics, does show distortion due to the decimation procedure. 
Such short distance distortions are typical of numerical decimation 
procedures. One is interested in extracting an effective action 
that is good at intermediate and long scales. 

After blocking, we need 
to assume some model of the effective action. We take 
a single plaquette action 
\begin{equation}
S=\sum_{j=1/2}^{N_r}\beta_j[1-\frac1{d_j}\chi_j(U_p)],\label{eq:ef_act}
\end{equation}
truncated at some high representation $N_r$ as our general form of the 
effective action. (As usual, in (\ref{eq:ef_act}) $U_p$ denotes the 
product of the bond variables along the boundary of the  plaquette $p$.)

The choice (\ref{eq:ef_act}) is motivated by some exact analytical results 
\cite{Tomboulis:2005zr}. There are decimation transformations of 
the `potential moving' type characterized by one or more free parameters 
which, after each blocking step, preserve a single plaquette action 
of the type (\ref{eq:ef_act})  
albeit with the full (infinite) set of representations. By appropriate choice 
of the decimation transformation parameter(s), the partition function 
obtained after a blocking step can be made to be either an upper or a lower 
bound on the original partition function.  
It is then possible to introduce a single parameter which,
at each decimation step, interpolates between the upper and lower bound, and 
hence has a value that keeps the partition function 
constant, i.e. exact under each successive decimation step. 
The same result can be obtained for `twisted' partition function 
(partition functions in the presence of external fluxes) and some other 
related long-distance quantities. This means that the action 
(\ref{eq:ef_act}) can in principle reproduce the exact partition function 
and other judiciously chosen quantities under successive blockings.

To compare the effective action model  to the decimated original theory, we 
need an efficient way to simulate a gauge theory with action 
(\ref{eq:ef_act}). We use a procedure due to
Hasenbusch and Necco \cite{Hasenbusch:2004yq}. The fundamental representation 
part of the action with specially tuned coupling is used to generate 
trial matrices for the metropolis updating. This procedure typically
achieves $80\%$ acceptance rate for the metropolis algorithm at the used 
couplings. Alternatively one could use a newly developed biased metropolis 
algorithm \cite{Bazavov:2005vr}. Simple heatbath updating is 
used only in the case of the action restricted to only the 
fundamental representation. 

To measure couplings we use the microcanonical evolution method 
\cite{Creutz:1983ra}. For the microcanonical updating and demon 
measurements we implement an improved algorithm given in \cite{Hasenbusch:1994ne}.  
The demons energies are restricted to $[-E_{max},$ $E_{max}]$, thus
preventing demons from `running away' with all the energy. We use $E_{max}=5$ value. 
The couplings 
$\beta_j$ of the effective action can be obtained as solutions of the equation
\begin{equation}
 \langle E^d_{j}\rangle = 1/\beta_j
 - E_{max}/tanh[\beta_j E_{max}], 
\end{equation}
where $E^d_{j}$ denotes the corresponding demon energy. In table 
(\ref{tab:dem_test}) we demonstrate the ability of the canonical demon
method in measuring the couplings on $8^4$ lattice. An ensemble of 
$3000$ configurations with couplings listed in the first row of the table is
used. Demon is allowed 1 sweep for reaching equilibrium, than 10 sweeps for
measurements. The measured couplings are listed on the second row of the table
and are in good agreement with the initial values.
\begin{table}[ht]
 \centering
 \begin{tabular}{|c|c|c|c|c|c|}
  \hline
  & $\beta_{1/2}$&$\beta_1$&$\beta_{3/2}$&$\beta_2$&$\beta_{5/2}$\\\hline
  in   &2.2578& -0.2201& 0.0898& -0.0333& 0.0125\\\hline
  demon&2.2580(4)&-0.2206(4)&0.0903(5)&-0.0336(5)& 0.0127(4) \\
  \hline
 \end{tabular}
 \caption{\label{tab:dem_test}Measurements of couplings by canonical 
demon method.}
\end{table}

\section{Decimation study}
\begin{figure}[ht]
 \includegraphics[width=0.9\textwidth]{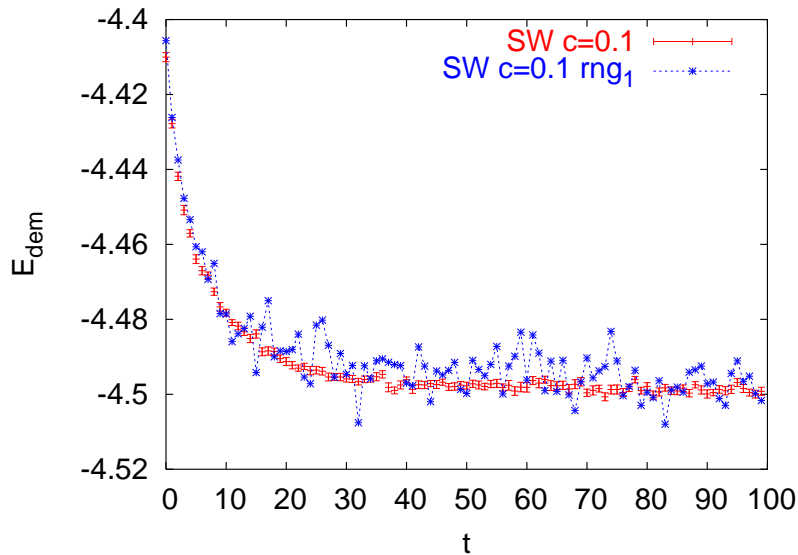}
 \caption{\label{fig:d_flow1}Demon fundamental representation 
energy flow for $c=0.1$ 
Swendsen decimated configurations. The average and a single demon run.
 }
\end{figure}

We fix the effective action to have 8 consecutive representations, starting
from the fundamental. For consecutive Monte Carlo updating of the effective 
action we truncated the number of couplings to the first five.
A  $32^4$ lattice at $\beta=2.5$ is decimated once,
using Swendsen type decimation with various staple weights $c$. In Fig.
\ref{fig:d_flow1} we show the fundamental representation demon energy flow,
starting from $c=0.1$ Swendsen decimated configurations. For 
the demon energy flow measurements we typically use from 20 to 100 replicas
(identical runs with different initial random number generator seeds).
\begin{figure}[ht]
 \includegraphics[width=0.9\textwidth]{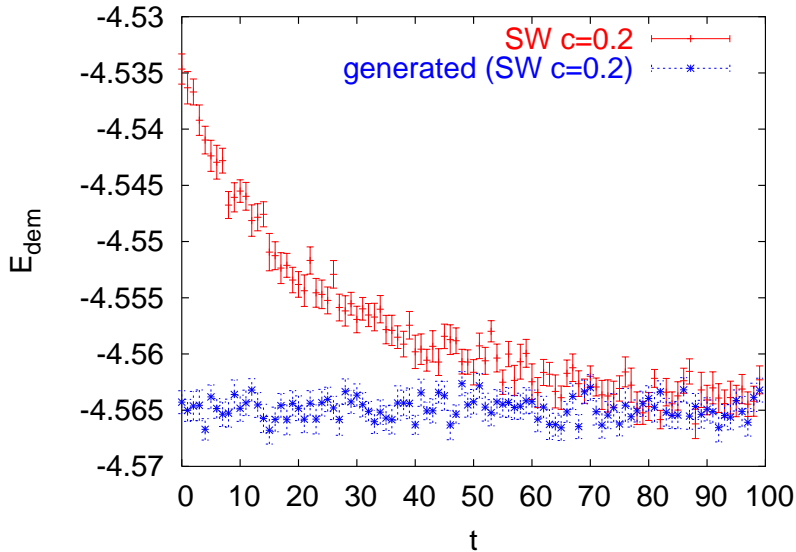}
 \caption{\label{fig:d_flow2}Demon fundamental representation 
energy flow for $c=0.2$ 
Swendsen decimated configurations and for configurations generated with 
an effective action.
 }
\end{figure}
The obvious main feature in this plot is that there is a significant demon 
energy change during microcanonical evolution. The change for different 
replicas is always in the same direction. There is a noticeable trend for flow 
stabilization at $\sim 100$ sweeps.

Next, taking $c=0.2$, we let the demon reach equilibrium ($>100$ sweeps) 
and then measure
the couplings of the effective action (\ref{eq:ef_act}). We then simulate
the effective action at these couplings and generate thermalized 
configurations for it. We then compare
the demon evolution on these thermalized configurations with the demon
evolution on the $c=0.2$ Swendsen decimated configurations (Fig 
\ref{fig:d_flow2}). There is now a striking difference. 
We see that in the case of the thermalized configurations of the effective 
action there is no change in the demon energy, which indicates a very
fast demon equilibration. Whereas in the case of the decimated configurations 
there is a pronounced energy change (as we also saw with the $c=0.1$ 
Swendsen decimated configurations). This pronounced energy change is 
clearly due to configuration 
equilibration during microcanonical evolution. This evolutions eventually 
brings  the original decimated configurations towards 
equilibrium configurations of the effective model.

The implication of this is obvious. 
Suppose one measures the couplings for the effective model from 
the decimated configurations after one or a few demon sweeps (i.e. 
on the configurations as obtained right after the decimation). These couplings 
are given in the first entry of the first column of Table \ref{tab:0_100}
below. Suppose one generates thermalized configurations of the 
effective action at {\it these} couplings. Then the decimated configurations 
are {\it not} representative of these effective action equilibrium 
configurations. 
As we saw the decimated configurations will evolve under 
microcanonical evolution towards equilibration at a set of different 
values for the couplings of the effective action. But by then they no longer 
are the true original decimated configurations obtained from the underlying 
finer lattice. 

This clearly would appear to present a potentially serious problem. 
It means that MC simulations using 
the effective action with coupling measured on the decimated configurations 
will not in general reproduce results from measurements obtained from 
the decimated configurations. 
Sufficient microcanonical evolution has to occur on the decimated 
configurations in order to `project' them into the equilibrium configurations 
of the effective model at {\it some} (other) set of couplings. It is 
an interesting question to what extent these evolved decimated configurations 
still retain any useful information concerning the starting action on 
the finer lattice. We come back to this point in subsection \ref{sec:observ} below.

The obvious next question is whether one can address this problem by 
fine-tuning the decimation procedure. 
Ideally, one would like to have for the measurement of couplings on the 
decimated configurations the same situation as that seen in the 
measurement of couplings on the undecimated configurations (cf. 
Table \ref{tab:dem_test} above), i.e. very fast demon thermalization 
indicating that the configurations are equilibrium configurations of 
the action for which the couplings are being measured.

\begin{figure}[p]
  \includegraphics[width=0.9\textwidth]{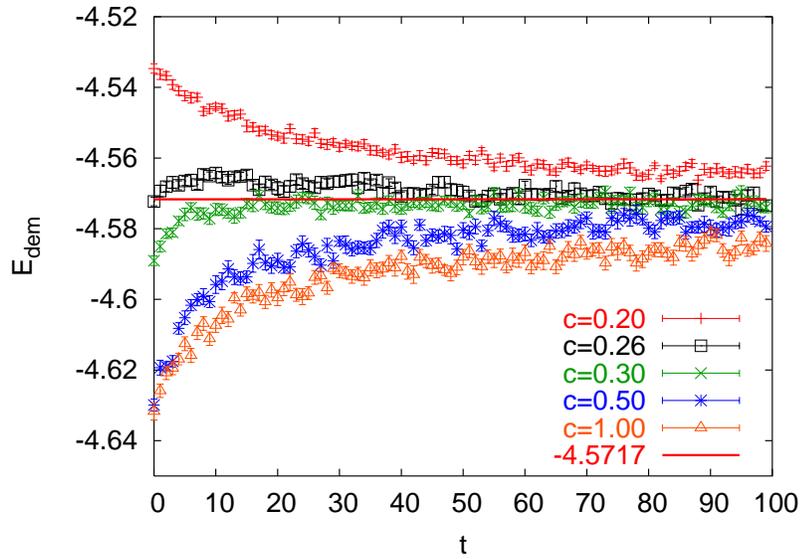}
  \caption{\label{fig:sw_dec_dem}Demon fundamental representation 
energy flow for  Swendsen decimation at various $c$ values.}
\end{figure}
\begin{figure}[p]
  \includegraphics[width=0.9\textwidth]{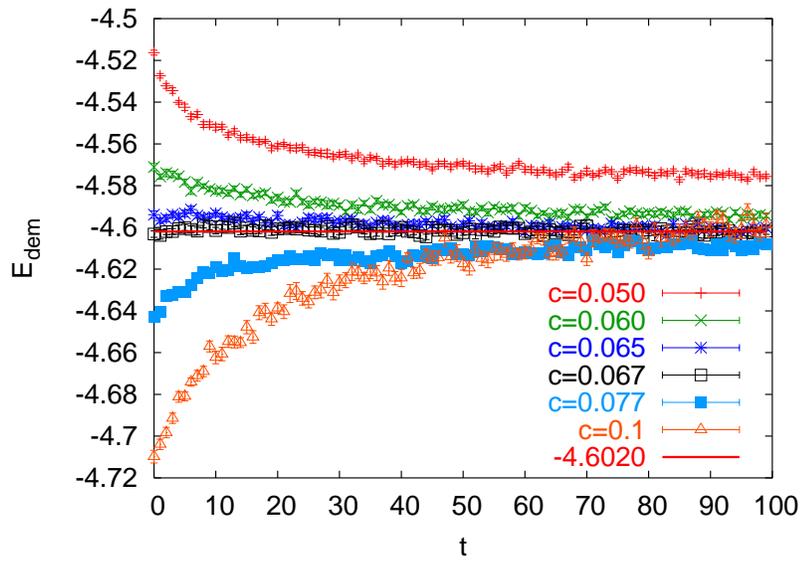}
  \caption{\label{fig:ds_dec_dem}Demon fundamental representation 
energy flow for DSB decimation at various $c$ values.}
\end{figure} 
The only freedom in the  
specification of the decimations (\ref{eq:sdec}), (\ref{eq:dsbdec}) 
is the staple weight parameter $c$.  
We then vary $c$ and observe the demon energy
flow. In Fig. \ref{fig:sw_dec_dem} we exhibit the fundamental demon energy
evolution for $c=0.2,\ldots,1.0$ Swendsen decimations.
We observe that there is a special $c\approx0.26$ value, when right from the
start there is little demon energy change. These particular decimation
configurations are then very close to the equilibrium configuration of 
the action (\ref{eq:ef_act}).  In Fig \ref{fig:ds_dec_dem} we 
show the fundamental demon energy flow for DSB 
decimations with $c=0.050, \ldots, 0.1$. Again, there is a special value 
in the vicinity $c\sim 0.067$ for which there is no significant 
flow from the outset. 

\begin{figure}[p]
  \includegraphics[width=0.9\textwidth]{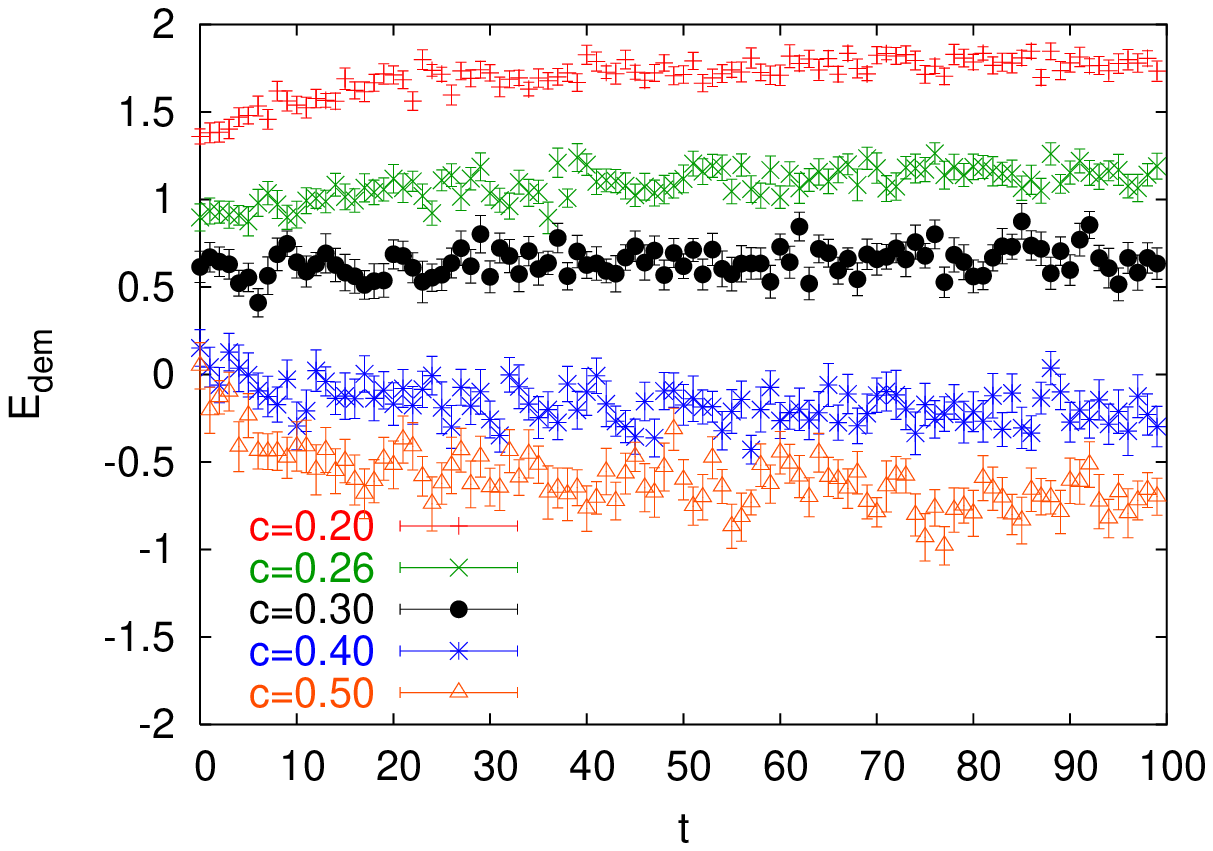}
  \caption{\label{fig:dem_sw_adj}Demon adjoint representation 
energy flow for Swendsen decimation at various $c$ values. }
\end{figure}

\begin{figure}[p]
  \includegraphics[width=0.9\textwidth]{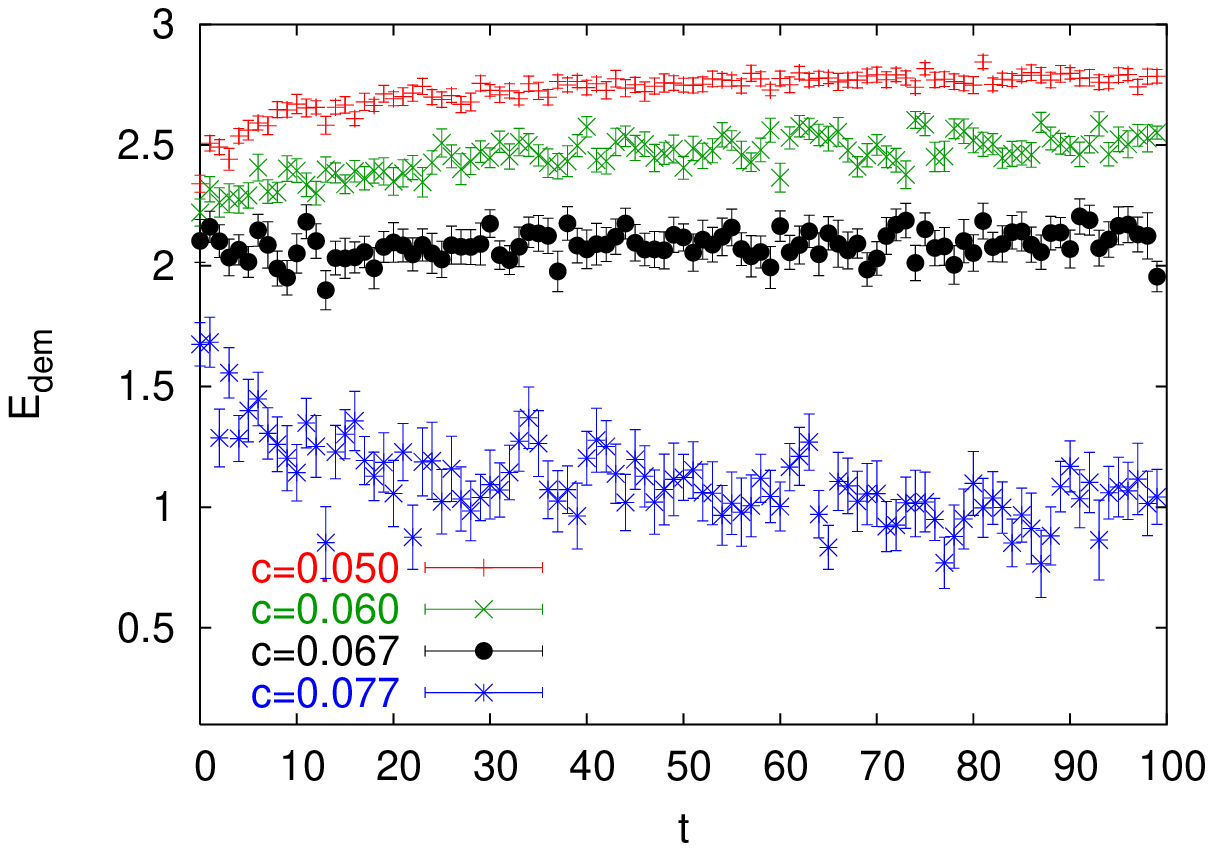}
  \caption{\label{fig:dem_ds_adj}Demon adjoint representation 
energy flow for DSB decimation at various $c$ values.}
\end{figure}

In Figs. \ref{fig:dem_sw_adj}, \ref{fig:dem_ds_adj} we look at 
the adjoint demon energy flows for Swendsen and DSB decimations, respectively. 
For Swendsen decimations we notice that now there is a small change 
for $c=0.26$, while no appreciable 
flow for $c=0.3$. For DSB decimations there appears to be 
no discernible shift in the vicinity of the optimal $c$ value.  
This is the first indication that DSB decimation is better suited for the 
effective action (\ref{eq:ef_act}).

\subsection{Observables\label{sec:observ}} 

We next compare some medium scale physical observables measured on 
the decimated configurations (right after decimation) and on 
configurations obtained from the effective action as described below. 

First, for $c=0.2$ Swendsen decimation, we take  the effective action with 
couplings obtained by demon measurements immediately after the decimation. 
We then compute $N\times N$
Wilson loops measured in two ways: (a) on the decimated configuration 
immediately after the decimation, denoted $W_{N\times N}^{dec}$; and (b)  
on configurations generated 
with this effective action, denoted $W_{N\times N}^{gen}$. The difference 
\begin{equation}
{\Delta W_{N\times N} \over W_{N\times N}^{dec}}=
\frac{W_{N\times N}^{gen}-W_{N\times N}^{dec}}{W_{N\times N}^{dec}} 
\label{eq:delta}
\end{equation} 
is displayed in the first row of Table \ref{tab:0_100}. 
The second row displays this difference when the effective action is now 
taken with couplings obtained by demon measurements after 100 sweeps, i.e. 
at the end of the microcanonical evolution shown in Fig. \ref{fig:d_flow2}. 
\begin{table}[ht]
 \centering
 \small
 \begin{tabular}{|c||c|c|c|c|}
  \hline
  $sws/m$ &$\beta_{1/2}, \beta_1, \beta_{3/2}, \ldots$ &
  $\Delta W_{1\times1}/ W^{dec}_{1\times1}$ & 
  $\Delta W_{2\times2}/ W^{dec}_{2\times2}$ &
  $\Delta W_{3\times3}/ W^{dec}_{3\times3}$\\
  \hline \hline
  0/1&2.1391(5),-0.1628(9), &&&\\
      &0.0637(11),-0.0250(1),&-0.0642(1)&-0.2832(5)&-0.7196(9)\\
      &0.0098(15)&&&\\\hline
  100/20& 2.2963(4),-0.2351(5),&&&\\
     & 0.0955(7),-0.0357(9),&-0.0045(1)&-0.0296(10)&-0.3912(20)\\
     & 0.0131(11),-0.0050(12)&&&\\\hline
  \end{tabular}
  \caption{Demon-measured couplings after $c=0.2$ Swendsen 
decimation, and difference of various size Wilson loops 
measured on decimated
  versus effective-action-generated configurations. Measurements performed 
  right after the decimation (measurement: 1 sweep) and after
  100 sweeps (measurements: 20 sweeps).\label{tab:0_100}}
 \end{table}

The table nicely illustrates the discussion above. One sees that 
measurements performed with the effective action having couplings obtained 
from the decimated configurations deviate from the values measured on the 
decimated configurations themselves (first row). Furthermore, 
the discrepancy grows substantially with increasing length scale, 
becoming large for the 
intermediate scale $3\times 3$ loop. This is in fact the worst possible 
outcome - it is at intermediate and long scales that the decimated 
configurations preserve the information on the undecimated lattice. 
But it is not unexpected 
since the decimated configurations are not 
equilibrium  configurations of the effective action. 
There is noticeable improvement, though still not near 
agreement, when couplings are obtained from 
microcanonically evolved decimated configurations, which are then 
equilibrium configurations of the resulting effective action (second row). 
This seems to imply that the microcanonically evolved decimated 
configurations, at least for these observables, retain some of 
information encoded in the original decimated configurations.

We next compute the difference (\ref{eq:delta}) for a range of $c$ values.  
In these computations we use 30 independent runs of 30 measurements each
for the $\beta$ measurements, and $20-30$ independent runs each of 400
measurements for the effective action simulations.
The effective action couplings are obtained after $100$ demon sweeps 
when thermalization is reached (cf. Figs. \ref{fig:d_flow1} and 
\ref{fig:d_flow2}). The results for Swendsen decimations 
and for DSB are presented in Table \ref{tab:fix_c} and Table 
\ref{tab:fix_c_dsb}, respectively.   
\begin{table}[ht]
 \centering
 \begin{tabular}{|c||c|c|c|}
 \hline
 $c$ &$\beta_{1/2}, \beta_1, \beta_{3/2}, \ldots$ &  
 $\Delta W_{2\times2}/ W^{dec}_{2\times2}$ & $\Delta W_{3\times3}/ W^{dec}_{3\times3}$\\
 \hline \hline
 0.0 & 1.1340(2),-0.1974(2),&&\\
     & 0.0531(3),-0.0162(4),&-0.8141(4)&-0.9876(39)\\
     & 0.0054(3),-0.0020(4) &&\\\hline
 0.1 & 1.9912(3),-0.3085(4),&&\\
     & 0.0990(4),-0.0362(6),&-0.4160(6)&-0.8899(11)\\
     & 0.0139(7),-0.0045(8) &&\\\hline
 0.2 & 2.2963(4),-0.2351(5),&&\\
     & 0.0955(7),-0.0357(9),&-0.0296(10)&-0.3912(20)\\
     & 0.0131(11),-0.0050(12)&&\\\hline
 0.26& 2.3351(7),-0.1449(10),&&\\
     &0.0766(12),-0.0279(13),& 0.1502(11)&0.0926(29)\\
     &0.0084(17)&&\\\hline
 0.3 & 2.3447(8),-0.0869(12),&&\\
     &0.0628(14),-0.0236(15),&0.2545(12)&0.4559(41)\\
     &0.0075(20)&&\\\hline
 0.4 & 2.3555(9), 0.0229(14),&&\\
     &0.0301(18),-0.0101(22),&0.4191(14)&1.1763(64)\\
     &0.0016(22)&&\\\hline
 0.5 & 2.3618(9),0.0866(13),&&\\
     & 0.0070(17),-0.0027(20),&0.4780(14)&1.5029(69)\\
     &-0.0013(22)&&\\\hline
 1.0 & 2.4033(9),0.1150(14),&&\\
     &-0.0274(18), 0.0071(22),&0.4456(14)&1.4845(75)\\
     &-0.0041(29)&&\\
\hline
\end{tabular}
\caption{Swendsen decimations. Demon-measured couplings at 
different $c$ values, and
difference of various size Wilson loops measured on decimated
versus effective-action-generated configurations. 
Thermalization: 100 sweeps, measurements: 20 sweeps.\label{tab:fix_c}.}
\end{table}
\begin{table}[ht]
 \centering
 \begin{tabular}{|c||c|c|c|c|}
 \hline
 $c$ &$\beta_{1/2}, \beta_1, \beta_{3/2}, \ldots$ & 
 $\Delta W_{2\times2}/ W^{dec}_{2\times2}$ & $\Delta W_{3\times3}/ W^{dec}_{3\times3}$&$\Delta W_{4\times4}/ W^{dec}_{4\times4}$ \\
 \hline \hline
 0.050&2.3536(5),-0.4208(9)&&&\\
      &0.1430(11),-0.0558(13)&-0.1817(6)&-0.637(1)&\\
      &0.0238(13),-0.0094(15)&&&\\\hline
 0.060&2.4660(7),-0.3635(11)&&&\\
      &0.1242(17),-0.0475(21)&0.0105(7)&-0.239(2)&\\
      &0.0195(25),-0.0070(24)&&&\\\hline
 0.063&2.4891(7),-0.3331(11)&&&\\
      &0.1140(14),-0.0436(19)&0.0800(9)&-0.049(3)&\\
      &0.0180(25),-0.0070(25)&&&\\\hline
 0.065&2.5023(7),-0.3098(12)&&&\\
      &0.1057(16), -0.0397(16)&0.1305(9)&0.106(3)&-0.034(14)\\
      &0.0145(14),-0.0029(15)&&&\\\hline
 0.067&2.5125(7),-0.2832(16)&&&\\
      &0.0964(25),-0.0367(29)&0.1774(9)&0.266(3)&0.290(19)\\
      &0.0139(29)&&&\\\hline 
 0.077& 2.5463(11),-0.1167(17),&0.4149(14)&1.270(7)&\\
      & 0.0320(23),-0.0055(28)&&&\\\hline
   0.1&2.4762(20),0.4191(37)&&&\\
      &-0.1231(40),0.0504(39)&0.6558(9)&2.627(6)&\\
      &-0.0191(53),0.0063(54)&&&\\
 \hline
 \end{tabular}
 \caption{\label{tab:fix_c_dsb} Same as Table \ref{tab:fix_c} 
for DSB decimations.}
\end{table}
Clearly, the $c$ values that give the best results, giving a difference 
(\ref{eq:delta}) that 
goes to zero at intermediate 
size Wilson loops, are precisely those in the vicinity of the values 
that produce decimated configurations which 
are closest to equilibrium configurations of the effective action. 
These are around $c\sim 2.6$ for Swendsen decimations. 
For DSB decimations the optimal value is   
between $c > 0.065$ and $c < 0.067$. 
This then provides a method of fixing $c$ in (\ref{eq:sdec}) and 
(\ref{eq:dsbdec}). 
It is interesting to note in particular that the 
classical $c$ value of DSB produces results which are
incapable of reproducing the physics at these scales correctly.

\subsection{Other $\beta$'s}
\begin{table}[ht]
	\centering
	\begin{tabular}{|c|c|c||c|c|}
		\hline
		$\beta$& 1/2 & 1 &1/2&1\\\hline\hline
		2.5&0.26&0.3&0.067&0.068\\
		2.8&0.21&0.22&0.059&0.060\\
		3.0&0.19&0.20&0.056&0.057\\
		\hline

	\end{tabular}
	\caption{The $c$ parameter for no demon energy flow in fundamental 
        (1/2) and adjoint (1) representations for Swendsen (left part) and 
	DSB (right part) decimations at different $\beta$'s. 
        \label{tab:c_betas}}
\end{table}
So far we have been working at $\beta=2.5$ (Wilson action) on the  
undecimated lattice. In Table \ref{tab:c_betas} we list the optimal 
$c$ values that result in no demon energy flow also 
for some other $\beta$ values. As expected, the optimal value depends on 
$\beta$. 

Determination of one optimal $c$ appears somewhat less sharp for 
Swendsen decimations than for DSB decimations. The latter appear better behaved 
and exhibit more consistency between higher representation demon
energy flow and the fundamental energy. Overall, DSB is the better suited 
decimation procedure for the effective action (\ref{eq:ef_act}).

\subsection{Double decimation}
\begin{figure}[p]
  \includegraphics[width=0.9\textwidth]{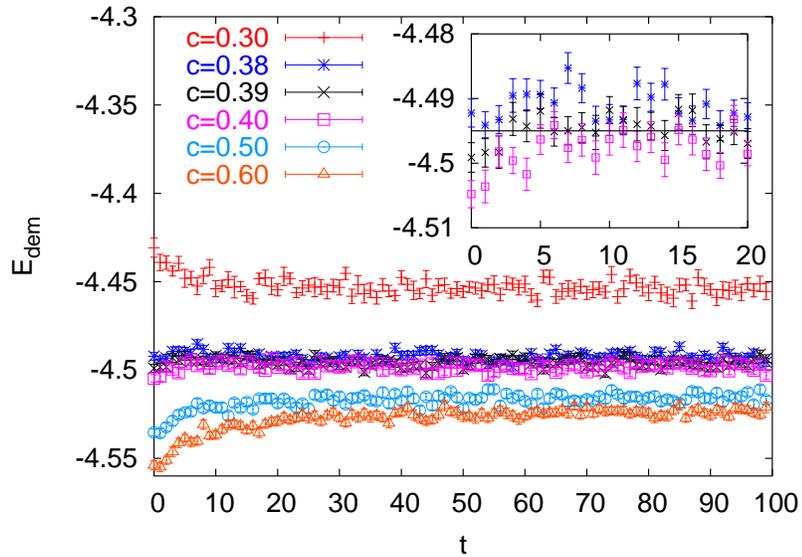}
  \caption{\label{fig:dem_sw_double}Fundamental representation 
demon energy flow for double Swendsen decimation at various $c$ values.}
\end{figure}

\begin{figure}[p]
  \includegraphics[width=0.9\textwidth]{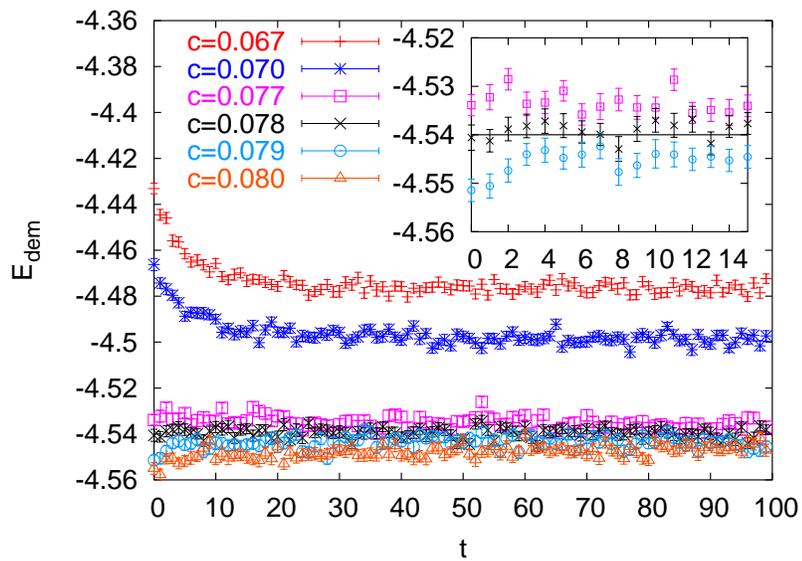}
  \caption{\label{fig:dem_ds_double}Fundamental representation demon 
energy flow for double DSB decimation at various $c$ values.}
\end{figure}

In Figs. \ref{fig:dem_sw_double} and \ref{fig:dem_ds_double} we show 
the fundamental representation demon energy flow after two successive Swendsen 
and DSB decimations, respectively, at various $c$ values at $\beta=2.5$.  
As seen in these plots, the general trend of demon energy flows  
is the same as after one decimation step. The optimal values for 
the staple weight $c$, 
however, differ from those for a single decimation step. 
For double decimation at $\beta=2.5$ we have optimal 
$c=0.39$ for Swendsen decimations; and $c=0.078$ for DSB decimations. 
At $\beta=2.8$, the optimal value for DSB decimations is $c=0.066$. 
%Thus, $c$ has, in general, to be readjusted for each successive decimation 
%step. See column 2 of Table \ref{tab:impr_act} below. 

\section{Improved action}
In this section we extract our improved action from our data. 
We have chosen DSB over Swendsen type decimation, since, as remarked above, 
it produces better overall results for the action (\ref{eq:ef_act}). 

\begin{figure}[ht]
  \includegraphics[width=0.9\textwidth]{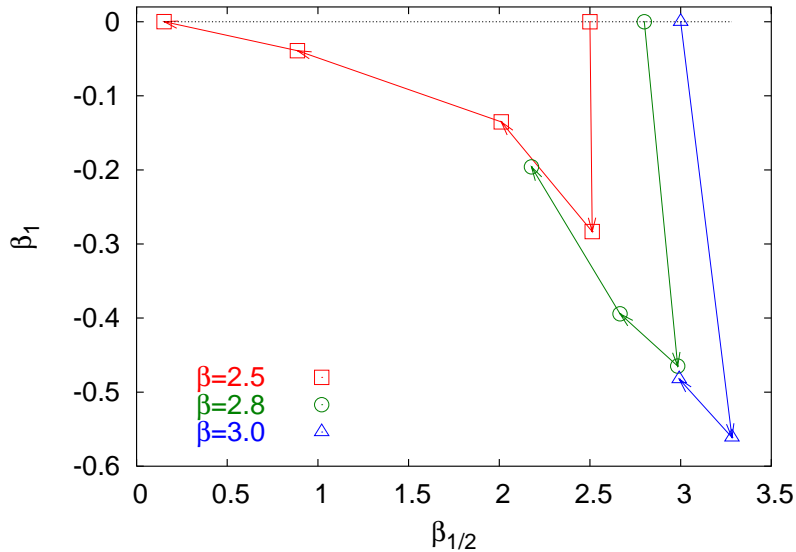}
  \caption{\label{fig:rg_flow} RG flow (from Table \ref{tab:impr_act}) 
under successive DSB decimations projected onto the $\beta_{1/2}$ - $\beta_1$ 
plane.}
\end{figure}

We look at 8 coupling in the improved action, and report
on first 5 of them. Typically we perform 30 independent runs each of $100-400$ 
measurements. We present the result in Table \ref{tab:impr_act}. 
At each successive 
decimation step, enumerated by $n$, we choose the value for
the staple weight which results in the minimal fundamental 
demon energy flow (cf. 4th column of Tab. \ref{tab:c_betas} for 
the $n=1$ step).
\begin{table}[ht]
 \centering 
 \begin{tabular}{|c|c||c|c|c|c|c|}
  \hline
  $n$ &$c$&$\beta_{1/2}$&$\beta_1$&$\beta_{3/2}$&$\beta_2$&$\beta_{5/2}$\\\hline\hline
  \multicolumn{7}{|c|}{$\beta=2.5$}\\\hline
  1&0.067&2.5125(7)&-0.2832(16)& 0.0964(25)&-0.0367(29)& 0.0139(29)\\
  2&0.078&2.0110(8)&-0.1351(7)&0.0385(10)&-0.0104(13)&0.0026(13)\\
  3&0.078*&0.8869(4)&-0.0390(4)&0.0067(4)&-0.0007(5)&0.0008(10)\\
  4&0.078*&0.1513(2)&0.0002(4)&-0.0003(5)&-0.0004(7)&0.0009(7)\\\hline\hline
  \multicolumn{7}{|c|}{$\beta=2.8$}\\\hline
  1&0.059&2.9841(23)&-0.4649(38)&0.1895(52)&-0.0898(62)&0.0432(65)\\
  2&0.066&2.6658(31)&-0.3943(49)&0.1446(61)&-0.0585(81)&0.0280(96)\\
  3&0.075&2.1773(16)&-0.1959(16)&0.0555(14)&-0.0164(16)&0.0057(23)\\
     \hline\hline
  \multicolumn{7}{|c|}{$\beta=3.0$}\\\hline
  1&0.056&3.2831(37)&-0.5611(53)&0.2397(74)&-0.1193(93)&0.05842(96)\\
  2&0.063&2.9920(54)&-0.4824(97)&0.181(14) &-0.067(18) &0.020(20)\\
  \hline
 \end{tabular}
 \caption{\label{tab:impr_act}Flow of couplings of the DSB decimated 
(improved) action. $\beta$ is the coupling of the original Wilson action 
on the original (undecimated) lattice. n enumerates 
successive decimations. First decimation is from 32 to 16 lattice, 
then all consecutive decimations are on 16. * means there is in fact 
virtually no discernible demon flow in reasonable range of $c$ around 
the indicated value.}
\end{table} 
In Fig. \ref{fig:rg_flow} we plot the RG flow of the first two, i.e. 
fundamental and adjoint, couplings from Table \ref{tab:impr_act}. 

\begin{table}[ht]
 \centering
 \begin{tabular}{|c||c|c|c|c|}
 \hline
 $c$ &$\beta_{1/2}, \beta_1, \beta_{3/2}, \ldots$ & 
 $\Delta W_{1\times1}/ W^{dec}_{1\times1}$ & 
 $\Delta W_{2\times2}/ W^{dec}_{2\times2}$ & $\Delta W_{3\times3}/ W^{dec}_{3\times3}$\\
 \hline \hline
0.067&2.5125(7),-0.2832(16)&&&\\
      &0.0964(25),-0.0367(29)&-0.0018(1)&0.1774(9)&0.266(3)\\
      &0.0139(29)&&&\\\hline 
0.067&2.4574(5),-0.1824(4)&-0.0012(1)&0.180(1)&0.273(4)\\
 \hline
 \end{tabular}
 \caption{Comparison of measurement of 5(8)
couplings with only first 2 couplings effective action. DSB decimation 
at $\beta=2.5$. The first row is taken from Table \ref{tab:fix_c_dsb}
\label{tab:impr_act_2a}above.} 
\end{table}
\begin{table}[ht]
 \centering
 \begin{tabular}{|c|c|c|}
  \hline
  $\beta$& $\beta_{1/2}$&$\beta_1$\\\hline
  2.5&2.4574(5)&-0.1824(4)\\
  2.8&2.8366(6)&-0.2428(4)\\
  3.0&3.0802(9)&-0.2693(6)\\
  \hline
 \end{tabular}
 \caption{\label{tab:impr_act_2b}Couplings of the decimated (improved) action  
 retaining only first 2 representations for DSB decimation 
at different $\beta$'s.}
\end{table}

As a matter of practical expediency, one may want to consider the effective 
action truncated to just these two couplings at the outset. In Table 
\ref{tab:impr_act_2a} we compare the results of measurements 
keeping just the two couplings to those keeping all five (eight). 
Comparison of the two rows of this table indicates the size of 
systematic error induced by this truncation of the effective action.  
The values of the two couplings for different starting $\beta$'s is given 
in Table \ref{tab:impr_act_2b} to be compared with those in Table 
\ref{tab:impr_act} ($n=1$).

\section{Summary and outlook}
We studied MCRG decimations in $SU(2)$ LGT employing either DSB or 
Swendsen decimations followed by a search for an effective action 
in the space of multi-representation single-plaquette 
actions with up to eight couplings using the demon method. 
Examination of the demon microcanonical evolution on 
the decimated configurations reveals the following general feature. 
Given the couplings of the effective action obtained from 
the decimated configurations, consider the equilibrium configurations 
of the effective action at these couplings. Then 
the decimated configurations are not, in general, representative of these 
equilibrium configurations. This means that simulations with the 
effective action at these couplings will not reproduce measurements 
of observables obtained from the decimated configurations as  
demonstrated in section 3 above.   

If sufficient microcanonical evolution of the decimated configurations 
is allowed, they will eventually result into configurations 
that are indeed equilibrium configurations of the effective action, that is 
the effective action at couplings obtained from these evolved decimated 
configurations. But the evolved decimated configurations are no longer 
the original decimated configurations, and cannot be relied upon to 
still adequately encode information from the original undecimated lattice. 

Solving this problem means having decimated configurations that are 
already equilibrium configurations of the adopted form of the effective 
action at the couplings obtained from the decimated configurations. 
This in general requires fine-tuning of the 
decimation and/or the effective action. 

In the case of the type of decimations and effective action adopted in this 
study, we saw that this fine-tuning could be achieved by fixing 
the value of the staple weight parameter $c$ in the specification 
of the decimation procedure, and retaining sufficient number of couplings.    
Also, this tuning works somewhat better for DSB decimations 
than Swendsen decimations.   
The result is the improved action presented in Table \ref{tab:impr_act} and 
Fig. \ref{fig:rg_flow}. Further improvements and refinements are presumably 
possible if more elaborate decimations involving more parameters are 
employed. 

Clearly, the general state of affairs described here 
holds  independently of the choice of 
decimation procedure and/or effective action. In this study we used the 
multi-representation single-plaquette action. Preliminary data 
with alternative actions, such as a multiloop fundamental representation 
action, reveal the same picture as expected. 

The use of the demon method for measuring couplings is also immaterial. 
The alternative Schwinger-Dyson (SD) method could be used. With this 
latter method, however, one does not have the option of microcanonically 
evolving the decimated configurations towards 
equilibration vis-a-vis the effective action,  
which is very informative and a nice advantage of the demon method. 
With SD a necessary test is to compare between: (a) the expectations, 
computed from the decimated configurations, of the 
operators occurring in the SD equation, which are then used in 
that equation to obtain the couplings; and 
(b) the same expectations computed  
with effective action equilibrium configurations generated at these 
couplings.\footnote{In this connection, we note in passing that 
commonly used checks in the literature of the above various 
MCRG procedures for obtaining effective actions are comparisons of the 
string tension. These, however, are rather uninformative. It is well known 
that the string tension, an asymptotic long distance quantity, is actually 
insensitive to the form of the action, e.g. \cite{Hasenbusch:2004yq}. 
} We hope to report results employing these alternative choices 
elsewhere. 

Extension of our study to $SU(3)$ to obtain the analog of the $SU(2)$ 
effective action arrived at here would also be worthwhile.

\section*{Acknowledgments}
We thank Academical Technology Services (UCLA) 
for computer support. This work was in part supported by 
NSF-PHY-0309362 and NSF-PHY-0555693.

\end{document}